# Impactos da Navegação Baseada em Performance nos Tempos de Voo da Aviação Comercial

João Basílio Tarelho Szenczuk
Alessandro V. M. Oliveira⭢

Instituto Tecnológico de Aeronáutica, São José dos Campos, Brasil

⭢ Autor correspondente. Instituto Tecnológico de Aeronáutica. Praça Marechal Eduardo Gomes, 50. 12.280-250 - São José dos Campos, SP - Brasil.
E-mail: alessandro@ita.br.

*Resumo*: Este trabalho apresenta uma análise da literatura recente que examina fatores que influenciam os tempos de voo no Brasil, com atenção especial ao impacto das implementações de uma nova tecnologia, a Navegação Baseada em Performance (Performance Based Navigation – PBN). Os procedimentos PBN começaram a ser implementados no espaço aéreo brasileiro em 2009 e representam um novo conceito de navegação aérea, com uso de satélites, que cria rotas de voo 3D, mais curtas e precisas, possivelmente reduzindo as distâncias voadas e consequentemente o tempo dos voos e atrasos. O tempo dos voos depende de diversos fatores, como o tamanho e complexidade dos aeroportos de origem e destino, as condições meteorológicas, os modelos das aeronaves e o quão carregado estão os voos. Portanto, para que o impacto dos novos procedimentos seja avaliado é preciso que a variação nesse conjunto de fatores seja levada em consideração. Os resultados encontrados na literatura sugerem que a Navegação Baseada em Performance teve impacto positivo, diminuindo o tempo dos voos em cerca de 1-2%, o que representa uma economia de dezenas de milhares de horas de voo entre o início das implementações em 2009 até o fim de 2018.

*Palavras-chave*: transporte aéreo, companhias aéreas, econometria.

## I. Introdução

A pontualidade é um fator fundamental para medir a eficiência e o nível de serviço das companhias aéreas e aeroportos. Além disso, os atrasos nos voos trazem um conjunto de consequências indesejadas para o transporte aéreo, como custos extras para as companhias aéreas e passageiros, e limitações no espaço aéreo e nos aeroportos.

De acordo com a FAA (2018), o custo para uma companhia aérea por uma hora de atraso varia de cerca de US$ 1.400 a US$ 4.500. Já o valor do tempo do passageiro varia de US$ 35 a US$ 63 por hora. Ball et al. (2010) estima o custo de atrasos nos voos domésticos para a economia dos EUA, que atingiu US$ 32,9 bilhões em 2007. Este valor inclui os custos para as empresas aéreas e passageiros, além do custo da demanda perdida e do impacto estimado no Produto Interno Bruto (PIB). Segundo a Eurocontrol (2019), o espaço aéreo europeu gerou um total de 19,1 milhões de minutos de atraso em rota em 2018, o que significa, em média, 1,73 minutos por voo (a meta da Eurocontrol é reduzir este valor para 0,5 minutos por voo). Os meios para mitigar esses custos incluem melhorias no gerenciamento de tráfego aéreo e sistemas de navegação, com novas tecnologias e conceitos operacionais.

Neste âmbito, a Organização Internacional de Aviação Civil (Internacional Civil Aviation Organization – ICAO), desenvolveu o Plano de Navegação Aérea Global (Global Air Navigation Plan – GANP), cuja última edição foi publicada em 2019. Este plano representa uma metodologia estratégica que aproveita as tecnologias existentes e antecipa desenvolvimentos futuros com base em objetivos operacionais acordados pelos países e indústria. As implementações tecnológicas previstas por este plano são organizadas em blocos de seis anos que não se sobrepõem, começando em 2013 e continuando até 2031 (e além, conforme o passar dos anos). Essa abordagem estruturada fornece uma base para estratégias de investimento sólidas e gera comprometimento dos países, fabricantes de equipamentos, operadores e prestadores de serviços. Além disso, as inovações de acordo com o GANP podem ser implementadas de forma flexível, conforme a necessidade de cada espaço aéreo, desde que seja obedecido um mínimo essencial para o fluxo harmonioso de tráfego. Uma das principais prioridades do GANP é a Navegação Baseada em Performance.

Nesse contexto, cada país tem seu programa de implementação dessas novas tecnologias, coordenado pelos respectivos provedores de serviço de navegação aérea. No Brasil, o Departamento de Controle do Espaço Aéreo (DECEA) comanda o programa SIRIUS. Na Europa e Estados Unidos existem o SESAR e o NEXTGEN, controlados pela EUROCONTROL e FAA, respectivamente.

## II. Desempenho de Operações Aéreas

O conceito PBN representa uma mudança, de operações baseadas em auxílios fixos no solo, para operações baseadas na performance de um conjunto de sistemas, com uso de satélites, que cria rotas de voo 3D, mais curtas e precisas. Eles se dividem em duas especificações: Navegação de Área (Area Navigation – RNAV) e Performance de Navegação Requerida (Required navigation performance – RNP). A diferença entre RNAV e RNP é que o RNP requer um sistema embarcado de monitoramento e alerta. Cada procedimento RNAV ou RNP recebe uma designação numeral, que representa a acuracidade da navegação lateral do sistema em milhas náuticas. Por exemplo, RNP 4 indica que o sistema deve permitir o voo com desvio máximo de 4 milhas náuticas. Os novos procedimentos requerem que os sistemas embarcados, os sistemas de gerenciamento de tráfego aéreo e a qualificação da tripulação atendam a uma performance exigida, de acordo com a designação de cada procedimento, de acuracidade, integridade, disponibilidade e continuidade, que são definidos a seguir:

- Acuracidade: capacidade de manter o voo dentro da designação do procedimento durante 95% do tempo;
- Integridade: capacidade de manter o voo dentro de 2 vezes o limite da designação do procedimento durante 99,999% do tempo;

- Disponibilidade: a probabilidade do sistema de navegação não embarcado usado de manter os requisitos de acuracidade e integridade durante a navegação;
- Continuidade: a probabilidade do sistema de navegação da aeronave de manter a acuracidade e integridade durante a navegação.

O PBN, portanto, não depende de uma tecnologia ou auxílio à navegação específico, mas de um conjunto destes que atenda aos requisitos necessários para uma operação. Estas tecnologias ou auxílios podem ser baseados em solo, no espaço ou na aeronave.

O conceito fundamental a ser compreendido para o escopo deste trabalho é que os procedimentos PBN possibilitam voos mais diretos, com trajetórias mais lineares, pois as rotas não precisam passar acima de auxílios à navegação fixos no solo, já que os novos sistemas são capazes de balizar novas trajetórias. A Figura 1 ilustra essa vantagem. Se a distância voada for menor, possivelmente, o tempo de voo também será.

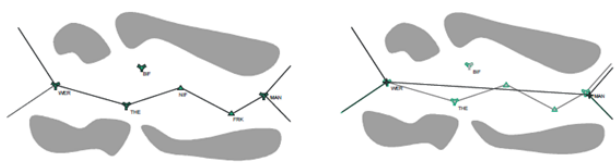

**Figura 1 – Rota convencional e rota PBN (ICAO, 2013)**

As implementações no Brasil foram iniciadas em 2009 pelo Departamento de Controle do Espaço Aéreo (DECEA) e são informadas ao público através das Circulares de Informação Aeronáutica (Aeronautical Information Circular – AICs). De acordo com as AICs os novos procedimentos foram implementados em 2009 nas terminais aéreas de Brasília, Recife, São Paulo e Rio de Janeiro, seguidas de Belo Horizonte em 2015 e toda a região sul do Brasil em 2017. Houve ainda duas outras fases de implementação nas terminais de São Paulo e Rio de Janeiro, em 2011 e 2013, respectivamente. A Figura 2 mostra as terminais equipadas com procedimentos PBN com círculos proporcionais ao tráfego aéreo de cada cidade.

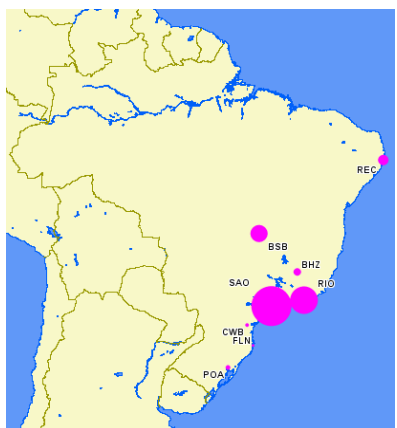

**Figura 2 – Áreas terminais com procedimentos PBN implementados**

Para se estimar o impacto dos procedimentos PBN nos tempos de voo é preciso levar em consideração a variação de todos os outros fatores que o influenciam, como características aeroportuárias, os tipos de aeronaves voando no espaço aéreo em consideração, as condições meteorológicas, o preço do combustível e características de mercado. Muitas pesquisas levam em consideração esses e outros fatores e aplicam técnicas econométricas para investigar as causas de atrasos em voo, conforme apresentado na seção a seguir.

## III. Revisão da Literatura

Duas vertentes principais de estudos econométricos de atrasos e tempos de voo podem ser identificadas na literatura, uma focada em fatores relacionados ao voo em si e outra nas características do mercado. Vejamos os principais trabalhos dessas duas vertentes a seguir.

A maioria dos estudos emprega proxies para possíveis determinantes de interrupções de voo, como condições climáticas e congestionamento no espaço aéreo e aeroportos. Hansen e Hsiao (2005), da University of California, Berkeley, estimam um modelo econométrico de atraso médio de chegada diária que incorpora os efeitos de filas de chegada, meteorologia, efeitos sazonais e efeitos seculares. Em Hansen & Hsiao (2006), filas de chegada em diferentes momentos do dia, a aviação geral, o tráfego militar e o volume de chegadas programadas (em oposição aos concluídos) foram adicionados ao modelo. Além disso, os autores sugerem que os impactos climáticos sobre os atrasos podem não ser determinados pelas condições meteorológicas em si, como foi assumido no primeiro modelo, mas pela interação entre o volume de chegadas programadas e as condições meteorológicas, como proposto no segundo trabalho.

Na outra vertente da literatura, preocupada com as características de mercado, Rupp, Owens e Plumly (2001), da East Carolina University, Texas A&M University e Valdosta State University, respectivamente, analisam a relação entre competição no nível de rotas e pontualidade, com dados do mercado norte-americano no período de 1997-2000. Os resultados indicam que rotas mais competitivas experimentaram mais atrasos nos voos, o que pode ser atribuído, segundo os autores, ao fato de que companhias adicionais em uma rota agrupam seus voos em horários de pico de demanda, gerando atrasos mais frequentes. Efeitos sazonais, restrições de capacidade nos aeroportos, demanda e efeitos de "hubbing" também foram fatores importantes. Mazzeo (2003), da Northwestern University, no entanto, sugere que a concentração de mercado está positivamente correlacionada com pior pontualidade, especialmente se apenas uma companhia aérea opera em uma rota. Outros fatores importantes para explicar os atrasos nos voos foram o clima, o congestionamento e as decisões de agendamento. Fica evidente, portanto, que esta é uma literatura ainda sem consenso.

Mayer & Sinai (2003), da Columbia University, também investigaram o mercado dos EUA, no período de 1988-2000. Seus resultados indicam que o efeito de "hubbing" é o principal contribuinte de congestionamentos. Além disso, as próprias companhias aéreas que operam no modelo hub-and-spoke incorrem todo o atraso causado pelos seus hubs. Santos e Robin (2010), da University of Oxford, seguem uma metodologia similar para o mercado europeu. Os fatores analisados foram efeitos idiossincráticos dos aeroportos e companhias aéreas, concentração em aeroportos, coordenação de slots, demanda e efeitos sazonais. Em oposição a Mayer & Sinai (2003), seus resultados indicam que embora os atrasos sejam maiores nos aeroportos hub, as empresas que operam no modelo hub-and-spoke experimentam nos aeroportos hub menores atrasos em comparação com as empresas que não operam no modelo hub-and-spoke.

Com relação aos impactos de uma otimização do tráfego aéreo especificamente, Guzhva, Abdelghany & Lipps (2014), da Embry-Riddle Aeronautical University, avaliam a implementação de um Sistema de Gerenciamento de Chegada de Aeronaves (Aircraft Arrival Management System - AAMS) com um modelo de regressão. O AAMS é parte do programa NextGen da FAA e é projetado para aumentar a eficiência das chegadas em aeroportos congestionados. O sistema pré-condiciona o fluxo de tráfego, fornecendo a cada aeronave um tempo requerido de chegada, à uma distância que permite aos pilotos um pequeno ajuste de velocidade (15 nós). O AAMS foi implementado em apenas 6,5% das chegadas de uma companhia aérea no Aeroporto Internacional Charlotte-Douglas (CLT) e essa porcentagem relativamente baixa de voos não foi suficiente para fornecer uma mudança perceptível no sistema aeroportuário como um todo. No entanto, os voos participantes de fato experimentaram tempos de permanência menores na área terminal.

## IV. PBN no Brasil

No Brasil, Szenczuk & Oliveira (2019) investigaram o impacto dos procedimentos PBN utilizando dados de 350 rotas domésticas envolvendo as capitais brasileiras no período entre janeiro de 2002 e dezembro de 2018. Levando em consideração os fatores apontados na literatura acima, o estudo emprega um modelo econométrico em que o tempo médio dos voos, incluindo tempos de táxi, em cada rota e mês, é a variável dependente, ou seja, a variável que se pretende explicar. Além do impacto dos novos procedimentos, o modelo emprega variáveis explicativas que permitem a análise dos principais fatores que causam variabilidade no tempo dos voos. Estes fatores podem ser divididos em quatro categorias: 1) fatores operacionais e custos, 2) características aeroportuárias, 3) competição de mercado, 4) características intrínsecas de cada mês e rota, como meses de férias escolares ou rotas com maior incidência de ventos.

A Tabela 1 mostra os sinais dos coeficientes que representam o efeito médio estimado do PBN do estudo dos autores, em percentual do tempo de voo e total de horas, para diferentes circunstâncias. Para as rotas com PBN na origem e no destino, ou seja, voos que ocorreram entre as cidades destacadas na Figura 2, o tempo dos voos foi reduzido em 0,48%, em média. Já as rotas com PBN apenas na origem apresentaram redução média de 0,84% enquanto as rotas com os novos procedimentos apenas no destino tiveram redução média de 1,29%. O restante da tabela mostra o efeito de cada fase de implementação nas terminais aéreas de São Paulo e Rio de Janeiro. Pode-se observar que a primeira fase teve efeito contrário ao esperado. Tal fato é possivelmente explicado pelas dificuldades operacionais enfrentadas durante as adaptações aos novos procedimentos. Na segunda fase, os resultados sugerem que apenas as rotas com destino a São Paulo ou Rio tiveram seus tempos de voo afetados, com redução média de 0,79%. Apenas após a terceira fase de implementação as rotas envolvendo as duas terminais tiveram a média dos tempos de voo reduzida em todos as circunstancias analisadas. Para as rotas envolvendo São Paulo ou Rio e outra capital com procedimentos PBN a redução média de tempo de voo atingida após a terceira fase foi de 0,97%. Para as rotas com origem em São Paulo ou Rio e destino a uma cidade sem os novos procedimentos, a redução média foi de 1,74%, enquanto as rotas com São Paulo ou Rio como destino tiveram uma redução média mais expressiva, de 3,33%.

Dentre os fatores operacionais e custos, foram considerados os efeitos da quantidade de passageiros na rota, a frequência da rota, o load factor, o preço do combustível, a presença de slots, a proporção de voos atrasados e a composição da frota. A Figura 4 mostra o resumo dos resultados, sendo: (+) para fatores que aumentam os tempos de voo, (-) para fatores que diminuem os tempos de voo, e (NS) para características estatisticamente não significantes.

Quanto maior o número de passageiros transportados em uma rota, maior tende ser o tempo dos voos, o que era esperado, já que aeroportos e espaços aéreos mais demandados tem mais congestionamentos. A frequência da rota, que neste caso é quantas vezes ela foi voada em um determinado mês, tende a reduzir o tempo dos voos. Os autores levantam algumas hipóteses que podem explicar isoladamente ou em conjunto este resultado: rotas mais voadas tem algum privilégio no gerenciamento de tráfego aéreo; companhias aéreas procuram voar mais rápido em rotas mais frequentes para gerenciamento de custos relacionados ao tempo de operação; companhias aéreas, pilotos e controladores estão mais acostumados ou adaptados às operações mais frequentes; as rotas mais voadas tem distâncias menores, o que não é explicitamente controlado por uma variável do modelo mas pelo método de estimação.

**Tabela 1 - Efeitos do PBN nos tempos de voo**

| Fator analisado | Efeito médio em % do tempo de voo | Efeito total em horas de voo |
|---|---|---|
| PBN na origem e destino | -0,48 | -7224 |
| PBN na origem | -0,84 | -27462 |
| PBN no destino | -1,29 | -42325 |
| 1ª fase São Paulo-Rio e Cidades com PBN | NS | NS |
| 2ª fase São Paulo-Rio e Cidades com PBN | NS | NS |
| 3ª fase São Paulo-Rio e Cidades com PBN | -0,97 | -24875 |
| 1ª fase São Paulo e Rio como origem | 0,84 | 25663 |
| 2ª fase São Paulo e Rio como origem | NS | NS |
| 3ª fase São Paulo e Rio como origem | -1,74 | -46679 |
| 1ª fase São Paulo e Rio como destino | 1,98 | 61604 |
| 2ª fase São Paulo e Rio como destino | -0,79 | -23467 |
| 3ª fase São Paulo e Rio como destino | -3,33 | -91219 |

*NS = efeito não significante.

Outro fator que interfere na velocidade dos voos é o load factor, que indica o quão carregado os aviões estavam. Aviões mais pesados precisam voar mais rápido para produzir a mesma força de sustentação, portanto, quanto maior o load factor, menor o tempo dos voos, controlando-se os demais fatores. Além disso, quanto mais passageiros estiverem em um voo atrasado, maior o custo do atraso para a companhia aérea. Um fator também incluso no modelo foi o tamanho médio das aeronaves, pelo mesmo motivo do load factor, já que aviões maiores também são mais pesados.

**Tabela 2 – Determinantes dos tempos de voo**

| Características | (1) Efeito nos tempos de voo |
|---|---|
| Quantidade de passageiros | + |
| Frequência de voos | − |
| Load Factor | − |
| Tamanho médio das aeronaves | − |
| Aeroporto controlado por slot | + |
| Preço do Cambustível | + |
| Proporção de voos atrasados na origem | − |
| Proporção de voos atrasados no destino | + |
| HHI da rota | − |
| HHI do aeroporto mais concentrado (destino ou origem) | − |
| Cidade com até 15 destinos | Caso base |
| Cidade com 15 a 44 destinos | + |
| Cidade com 45 a 69 destinos | + |
| Cidade com mais de 70 destinos | + |
| Menos de 10% de passageiros em conexão | Caso base |
| Entre 10 e 20% dos passageiros em conexão | + |
| Entre 20 e 30% dos passageiros em conexão | + |
| Mais de 30% dos passageiros em conexão | + |

*NS = efeito não significante.

Outro fator que pode aumentar o custo de um atraso e incentivar companhias a voarem mais rápido é se o aeroporto de destino é controlado por slots ou não. Slots são intervalos de tempo pré-determinados que uma companhia pode ocupar determinado portão de embarque em um aeroporto. Se um voo ultrapassa esse intervalo de tempo e fica incapacitado de embarcar ou desembarcar passageiros, o atraso pode se propagar para outros voos. Outro fator incluso no modelo é a proporção de voos atrasados na origem e no destino. Os resultados indicam que as companhias aéreas tentam recuperar um atraso na partida com maiores velocidades de voo e também que atrasos ocorrem em voo ou táxi nos aeroportos de destino.

O preço do combustível apresentou relação positiva com o tempo dos voos, como esperado. Quando o preço do querosene de aviação está alto, as empresas aéreas podem optar por voar em velocidades menores, consumindo menos combustível, mas gastando mais com salários e manutenção das aeronaves, que dependem do tempo de operação. Se os preços estão baixos, podem fazer o oposto.

Para estimar o efeito da complexidade do tráfego aéreo nos aeroportos foi adotada uma categorização. São quatro categorias: 1) cidades com aeroportos servindo até 14 destinos; 2) entre 15 e 44 destinos; 3) entre 45 e 70 destinos; e 4) acima de 70 destinos. Os resultados indicam que o tempo dos voos tende a aumentar com o tamanho do aeroporto, mas que não há diferença entre os dois grupos intermediários. Apesar dessas variáveis serem razoáveis para estimar o impacto do tamanho e complexidade dos aeroportos, elas não controlam os efeitos de "hubbing". Empresas que operam no modelo hub and spoke tendem a concentrar seus voos em pequenos intervalos de tempo a fim de atender o máximo de conexões possíveis. Entretanto, esta prática pode causar congestionamentos em períodos de pico. Para controlar este efeito, o estudo adota uma segunda categorização, que se refere à proporção de passageiros em conexão. Novamente, há quatro categorias: 1) até 10% de passageiros em conexão; 2) entre 10 e 20% em conexão; 3) entre 20 e 30%; e 4) acima de 30% dos passageiros em conexão. Os resultados sugerem que quanto maior a proporção de passageiros em conexão, maiores são os tempos de voo, fornecendo evidências de que há efeitos de "hubbing".

Conforme discutido nos estudos internacionais apresentados, a concentração de mercado nas rotas e nos aeroportos pode afetar o tempo dos voos. Por isso, o estudo adota uma variável HHI de concentração na rota e outra para concentração no aeroporto. Os resultados indicam que a concentração de mercado na rota e no aeroporto tendem a diminuir o tempo dos voos, corroborando os resultados de Mayer e Sinai (2003), Santos e Robin (2010) e Rupp et al. (2001).

Por fim, existem os fatores não observáveis, ou não explicitamente incluídos no modelo de regressão, que são particulares de cada mês ou rota, e que influenciam no tempo dos voos. No caso das rotas, por exemplo, há o efeito da distância de cada uma delas e da predominância de ventos. No caso dos meses, há o impacto de acontecimentos específicos em cada mês, como, por exemplo, más condições meteorológicas ou um apagão aéreo como ocorreu em meados da primeira década deste século. Esses fatores são controlados pelo método de estimação do modelo adotado.

## V. Conclusões

Este estudo revisa a literatura sobre fatores que afetam os tempos de voo na aviação comercial, destacando o papel da Navegação Baseada em Performance (PBN), implementada no Brasil a partir de 2009. A PBN usa satélites para definir trajetórias de voo mais diretas e precisas, visando diminuir distâncias percorridas e, por consequência, os tempos de voo e atrasos. A literatura considera variáveis como características dos aeroportos, condições climáticas, tipos de aeronaves e volume de tráfego.

Afinal, é possível encurtar o tempo de voo entre dois pontos através de uma modernização do espaço aéreo? Sim. Apesar da redução no tempo médio dos voos ser aparentemente modesta, em torno de 2%, isso corresponde a uma economia total de algumas dezenas de milhares de horas de voo desde o início das implementações até o final de 2018, o que pode representar uma redução significativa de custos operacionais diretos das companhias aéreas, como combustível, salários da tripulação, manutenção das aeronaves, entre outros.

Além disso, o modelo desenvolvido no estudo nacional revelou os principais fatores que explicam a variabilidade nos tempos de voo, como a demanda e a frequência das rotas, o tamanho dos aeroportos, a concentração de mercado e os potenciais custos dos atrasos. Os sugerem mostram que rotas e aeroportos mais concentrados têm voos mais rápidos. Além disso, os voos demoram mais em aeroportos maiores, o que é associado ao aumento da complexidade do gerenciamento de tráfego aéreo e das distâncias percorridas em solo nos principais aeroportos.